\documentclass[preprint,3p]{elsarticle}

\usepackage{amssymb}
\usepackage{amsthm}
\usepackage[mathlines]{lineno}
\usepackage{graphicx}
\usepackage{units}
\usepackage{url}
\usepackage{ulem}
\usepackage{amsmath}
\usepackage{amsfonts}
\usepackage{bm}
\usepackage{textcomp}
\usepackage{subfigure}
\usepackage{multicol}
\usepackage{verbatim}
\usepackage{rotating}
\usepackage[colorlinks,linkcolor=red,citecolor=red]{hyperref}
\usepackage{float}
\usepackage{epsfig}
\usepackage{dcolumn}
\usepackage{bm}
\usepackage{color}
\usepackage{pstricks}
\usepackage{pst-node}
\usepackage{times}
\usepackage{indentfirst}
\usepackage[english]{babel}
\usepackage{epstopdf}
\addto{\captionsenglish}{%

}
\journal{Physics Letters B}
\makeatletter
\newenvironment{tablehere}
  {\def\@captype{table}}
  {}

\newenvironment{figurehere}
  {\def\@captype{figure}}
  {}
\makeatother

\newcommand{\XXN}{\Xi^{0}\bar\Xi^{0}}
\newcommand{\XXB}{\Xi^{-}\bar\Xi^{+}}

\newcommand{\EE}{e^+e^-}
\newcommand{\BB}{B\bar{B}}

\newcommand{\jpsi}{J/\psi}
\newcommand{\ar}{\rightarrow}
\newcommand{\llb}{\Lambda\bar{\Lambda}}

\newcommand{\bfg}{\begin{figure}}
\newcommand{\efg}{\end{figure}}
\newcommand{\bitm}{\begin{itemize}}
\newcommand{\eitm}{\end{itemize}}
\newcommand{\bnum}{\begin{enumerate}}
\newcommand{\enum}{\end{enumerate}}
\newcommand{\btbl}{\begin{table*}}
\newcommand{\etbl}{\end{table*}}
\newcommand{\btbu}{\begin{tabular}}
\newcommand{\etbu}{\end{tabular}}
\newcommand{\bcl}{\begin{center}}
\newcommand{\ecl}{\end{center}}
\newcommand{\bbt}{\bibitem}
\newcommand{\beq}{\begin{equation}}
\newcommand{\eeq}{\end{equation}}
\newcommand{\beqr}{\begin{eqnarray}}
\newcommand{\eeqr}{\end{eqnarray}}

\begin{document}
\begin{frontmatter}
\title{\bf \boldmath  Measurement of cross section for $\EE\ar\XXN$ near threshold}
\author{
\small 
M.~Ablikim$^{1}$, M.~N.~Achasov$^{10,c}$, P.~Adlarson$^{67}$, S. ~Ahmed$^{15}$, M.~Albrecht$^{4}$, R.~Aliberti$^{28}$, A.~Amoroso$^{66A,66C}$, M.~R.~An$^{32}$, Q.~An$^{63,49}$, X.~H.~Bai$^{57}$
% ...
\\
\vspace{0.2cm}
(BESIII Collaboration)\\
\vspace{0.2cm} 
$^{1}$ Institute of High Energy Physics, Beijing 100049, People's Republic of China\\
% ...
$^{4}$ Bochum Ruhr-University, D-44780 Bochum, Germany\\
% ...
$^{10}$ G.I. Budker Institute of Nuclear Physics SB RAS (BINP), Novosibirsk 630090, Russia\\
% ...
$^{15}$ Helmholtz Institute Mainz, Staudinger Weg 18, D-55099 Mainz, Germany\\
% ...
$^{28}$ Johannes Gutenberg University of Mainz, Johann-Joachim-Becher-Weg 45, D-55099 Mainz, Germany\\
% ...
$^{32}$ Liaoning Normal University, Dalian 116029, People's Republic of China\\
% ...
$^{49}$ State Key Laboratory of Particle Detection and Electronics, Beijing 100049, Hefei 230026, People's Republic of China\\
% ...
$^{57}$ University of Jinan, Jinan 250022, People's Republic of China\\
% ...
$^{63}$ University of Science and Technology of China, Hefei 230026, People's Republic of China\\
% ...
$^{66}$ University of Turin and INFN, (A)University of Turin, I-10125, Turin, Italy; (B)University of Eastern Piedmont, I-15121, Alessandria, Italy; (C)INFN, I-10125, Turin, Italy\\
$^{67}$ Uppsala University, Box 516, SE-75120 Uppsala, Sweden\\
% ...
$^{c}$ Also at the Novosibirsk State University, Novosibirsk, 630090, Russia\\
% ...
\vspace{0.4cm}
}
\date{\today}
\begin{abstract}
Using  $e^+e^-$ collision data at ten center-of-mass energies between 2.644 and 3.080 GeV collected with the BESIII detector at BEPCII and corresponding to an integrated luminosity of about 500 pb$^{-1}$, we measure 
the cross sections and effective form factors 
 for the process $\EE\ar\XXN$  utilizing a single-tag method. 
A fit to the cross section of  $\EE\ar\XXN$ with a pQCD-driven power function is performed, from which no significant resonance or threshold enhancement  is observed.
In addition, the ratio of cross sections for the processes $\EE\ar\XXB$ and $\XXN$ is calculated using recent BESIII measurement and  is found to be compatible with expectation from isospin symmetry.
\end{abstract}
\end{frontmatter}

\begin{multicols}{2}
\section{Introduction}
In the past few decades, many experiments have observed surprising behavior in the near-threshold region in the production cross section of nucleon pairs in $\EE$ collisions. The measured cross section for the process $e^+e^-\to p\bar{p}$  is approximately constant in the energy range from threshold up to about 2 GeV~\cite{Delcourt:1979ed, Bisello:1983at, Antonelli:1993vz, ppbar, ppbar02, Akhmetshin:2015ifg, Pedlar:2005sj, ppbar03}, with an average value of  about 0.85~nb. 
Similar behavior in the near threshold region has been observed in the processes $\EE\to n\bar{n}$~\cite{nnbar, Ablikim:2021eqa, Druzhinin:2019gpo}, $\EE\ar\Lambda\bar\Lambda$~\cite{LLbar}  and $\EE\ar\Lambda^{+}_{c}\bar\Lambda^{-}_{c}$~\cite{Pakhlova:2008vn, LcLcbar}  with average cross sections of 0.8~nb, 0.3~nb and 0.2~nb, respectively.
The non-vanishing cross section near threshold and the wide-range plateau discussed above
have attracted much interest and driven many  theoretical studies, including scenarios that invoke $B\bar{B}$ bound states~\cite{theory1} or unobserved meson resonances~\cite{theory11}, Coulomb final-state interactions or quark electromagnetic interaction that into account the asymmetry between attractive and repulsive Coulomb factors~\cite{theory2, theory22}. 
In the present context of QCD and of our understanding
 of the quark-gluon structure of hadrons, it is particularly interesting to understand these anomalous phenomena in the hyperon sector~\cite{Aubert, DM2, BESIII_hyperon}.
Recently, the BESIII collaboration performed high precision studies of possible threshold enhancement in the  processes $\EE\ar\Sigma^{\pm}\bar\Sigma^{\mp}$~\cite{Ablikim:2020kqp} and $\XXB$~\cite{Ablikim:2020rwi} with an energy scan method, and also reported a non-vanishing cross section near threshold.
To understand the nature of these threshold effects,  measurements of the near-threshold pair production of other hyperons will be valuable.

Additionally, the electromagnetic structure of hadrons, parametrized in terms of electromagnetic form factors (EMFFs)~\cite{Denig:2012by}, provides a key to understanding QCD effects in bound states. 
However, experimental information regarding the EMFFs of hyperons remains limited.
The access to hyperon structure by EMFFs provides  extra motivation for measurements of exclusive cross sections and EMFFs for baryon antibaryon pairs.

In this Letter, we report a measurement of the Born cross section and the effective form factor for the process  $\EE\ar\XXN$ using a single-baryon-tag method
at center-of-mass (CM) energies between 2.644 and 3.080 GeV. We fit the Born cross sections under various hypotheses for the $\XXN$ production in $\EE$ annihilation.
The data set used in this analysis corresponds to a total of about 500 pb$^{-1}$ $e^+e^-$ collision data~\cite{luminosity, Ablikim:2017wlt} collected with the BESIII detector~\cite{besiii} at 
the BEPCII storage rings~\cite{bepcii}.

\section{BESIII detector and Monte Carlo simulation}
\label{sec:detector}
The BESIII detector~\cite{besiii} records symmetric $e^+e^-$ collisions 
provided by the BEPCII storage rings~\cite{bepcii}.
BESIII has collected large data samples in the $\tau$-charm threshold region~\cite{Ablikim:2019hff}. The cylindrical core of the BESIII detector covers 93\% of the full solid angle and consists of a helium-based
 multilayer drift chamber~(MDC), a plastic scintillator time-of-flight
system~(TOF), and a CsI(Tl) electromagnetic calorimeter~(EMC),
which are all enclosed in a superconducting solenoidal magnet
providing a 1.0~T magnetic field. The solenoid is supported by an
octagonal flux-return yoke with resistive plate counter muon-identification modules interleaved with steel. 
The charged-particle momentum resolution at $1~{\rm GeV}/c$ is
$0.5\%$, and the d$E$/d$x$ resolution is $6\%$ for electrons
from Bhabha scattering. The EMC measures photon energies with a
resolution of $2.5\%$ ($5\%$) at $1$~GeV in the barrel (end-cap)
region. The time resolution in the TOF barrel region is 68~ps. The end-cap TOF
system was upgraded in 2015 using multi-gap resistive plate chamber
technology, providing a time resolution of
60~ps~\cite{etof}.

To determine the detection efficiency 
100,000 simulated $\EE\ar\XXN$ events are generated for each energy point with phase space modeling in
 the \textsc{conexc} generator~\cite{conexc}, which takes into account the beam-energy spread and corrections from initial-states radiation (ISR).  
The $\Xi^{0}$ is simulated via \textsc{evtgen}~\cite{evtgen} in its decay to
the $\pi^{0}\Lambda$ mode with the subsequent decay $\Lambda\to
p\pi^{-}$ with phase space model. The anti-baryons are
allowed to decay inclusively according to the branching fractions from
the Particle Data Group (PDG)~\cite{PDG2020}
(unless otherwise noted, the charge-conjugate state of the $\Xi^{0}$ mode is included by default below).
The response of the BESIII detector is modeled with Monte Carlo (MC) simulations
using a framework based on \textsc{geant}{\footnotesize 4}~\cite{geant4}.
 Large simulated samples of generic $\EE \to
\text{hadrons}$ events (inclusive MC) implemented
by  the \textsc{conexc} generator~\cite{conexc}
are used to estimate background.

\section{Event selection}
\label{sec:evt_sel}
The selection of $\EE\ar\XXN$ events with a full reconstruction method
suffers from a  low reconstruction efficiency.  Hence, a
single baryon $\Xi^{0}$ tag technique is employed, {\it i.e.}, only one
$\Xi^{0}$ baryon is reconstructed via the $\pi^{0}\Lambda$ decay mode with
$\Lambda\ar p\pi^{-}$ and $\pi^{0}\ar\gamma\gamma$, and the presence of the anti-baryon $\bar\Xi^{0}$ is inferred
from the recoil mass of the detected $\Xi^{0}$.

Charged tracks are required to be reconstructed in the MDC within the angular coverage
of the MDC: $|\!\cos\theta|<0.93$, where $\theta$ is the polar angle
with respect to the $e^{+}$ beam direction.  Information from d$E$/d$x$
measured in the MDC combined with
the TOF is used to construct a  particle-identification probability 
for the hypotheses of a pion, kaon, and proton.
Each track is assigned to the particle type with the highest probability. Events with at least one negatively charged pion and one proton
are kept for further analysis.

Photons are reconstructed from isolated showers in the EMC. The energy deposited
in the nearby TOF counter is included to improve the reconstruction
efficiency and energy resolution. Photon energies are required to be
greater than 25 MeV in the EMC barrel region $|\!\cos\theta|<0.8$),
 or
greater than 50 MeV in the EMC end cap ($0.86 < |\!\cos\theta| < 0.92$). The showers in the angular range between
the barrel and the end cap are poorly reconstructed and are excluded from
the analysis. Furthermore, the EMC timing of the photon candidate must be
in coincidence with collision events, $0\leq t\leq 700$ ns, in order to suppress
electronic noise and energy deposits unrelated to the collision events. At least two photon candidates are required. 

To reconstruct the $\pi^{0}$ candidates, a one-constraint (1C) kinematic fit is employed for all $\gamma\gamma$ combinations by constraining the invariant mass of two photons to the $\pi^0$ nominal mass~\cite{PDG2020}, and  a requirement is placed on the goodness-of-fit, $\chi^{2}_{1C} < 20$, to suppress non-$\pi^0$ backgrounds.
Here the fitted momentum of $\pi^{0}$ is used to further analysis.
To reconstruct $\Lambda$ candidates, a secondary vertex fit~\cite{XUM} is
applied to all $p\pi^{-}$ combinations and those with a goodness-of-fit 
$\chi^{2} < 500$ are selected.  In the case when there is more than one combination satisfying this requirement then that one with  the minimum value of $|M_{p\pi}-m_{\Lambda}|$ among all $p\pi$ combinations is chosen,
where $M_{p\pi}$ is the invariant mass of the $p\pi$ pair, and $m_{\Lambda}$ is the nominal mass of the $\Lambda$ baryon~\cite{PDG2020}.
The $p\pi^{-}$
invariant mass  of the selected candidate is required to be within 5 MeV/$c^{2}$ of the nominal
$\Lambda$ mass, a criterion determined by optimizing the figure of merit (FOM)
$\frac{S}{\sqrt{S + B}}$ based on the MC simulation. Here $S$ is the
number of signal MC events and $B$ is the number of the background events expected from simulation.
To further suppress background from non-$\Lambda$ events, the $\Lambda$ decay
length,~{\it i.e.},, the distance between its
production and decay positions, is required to be greater than zero.
The $\Xi^{0}$ candidates are reconstructed from the combination of the selected $\pi^{0}$ and $\Lambda$ candidates by minimizing the variable $|M_{\pi^{0}\Lambda}-m_{\Xi^{0}}|$, where $M_{\pi^0\Lambda}$ is the invariant mass of the $\pi^0\Lambda$ pair, and $m_{\Xi^{0}}$ is the nominal mass of the $\Xi^{0}$ baryon taken from the PDG~\cite{PDG2020}.
In order to suppress background further, the $\pi^{0}\Lambda$ invariant mass is required to be within 10 MeV/$c^{2}$ of the nominal $\Xi^{0}$ mass, where the cut value is again set from FOM studies.

To select $\bar\Xi^{0}$ anti-baryon candidates, we use the
distribution of mass recoiling against the selected $\pi^{0}\Lambda$
system,
\begin{equation}
  M^{\rm recoil}_{\pi^{0}\Lambda} = \sqrt{(\sqrt{s}-E_{\pi^{0}\Lambda})^{2} - |\vec{p}_{\pi^{0}\Lambda}|^{2}},
\end{equation}
where $E_{\pi^{0}\Lambda}$ and $\vec{p}_{\pi^{0}\Lambda}$ are the
energy and momentum of the selected $\pi^{0}\Lambda$ candidate in the
CM system, and $\sqrt{s}$ is the CM energy.  
Figure~\ref{scatterplot} shows the distribution of
$M_{\pi^{0}\Lambda}$ versus $M^{\rm recoil}_{\pi^{0}\Lambda}$ for all energy points. A clear accumulation is observed around the $\Xi^0$ nominal mass.
\begin{figurehere}
\bcl
\includegraphics[width=0.40\textwidth]{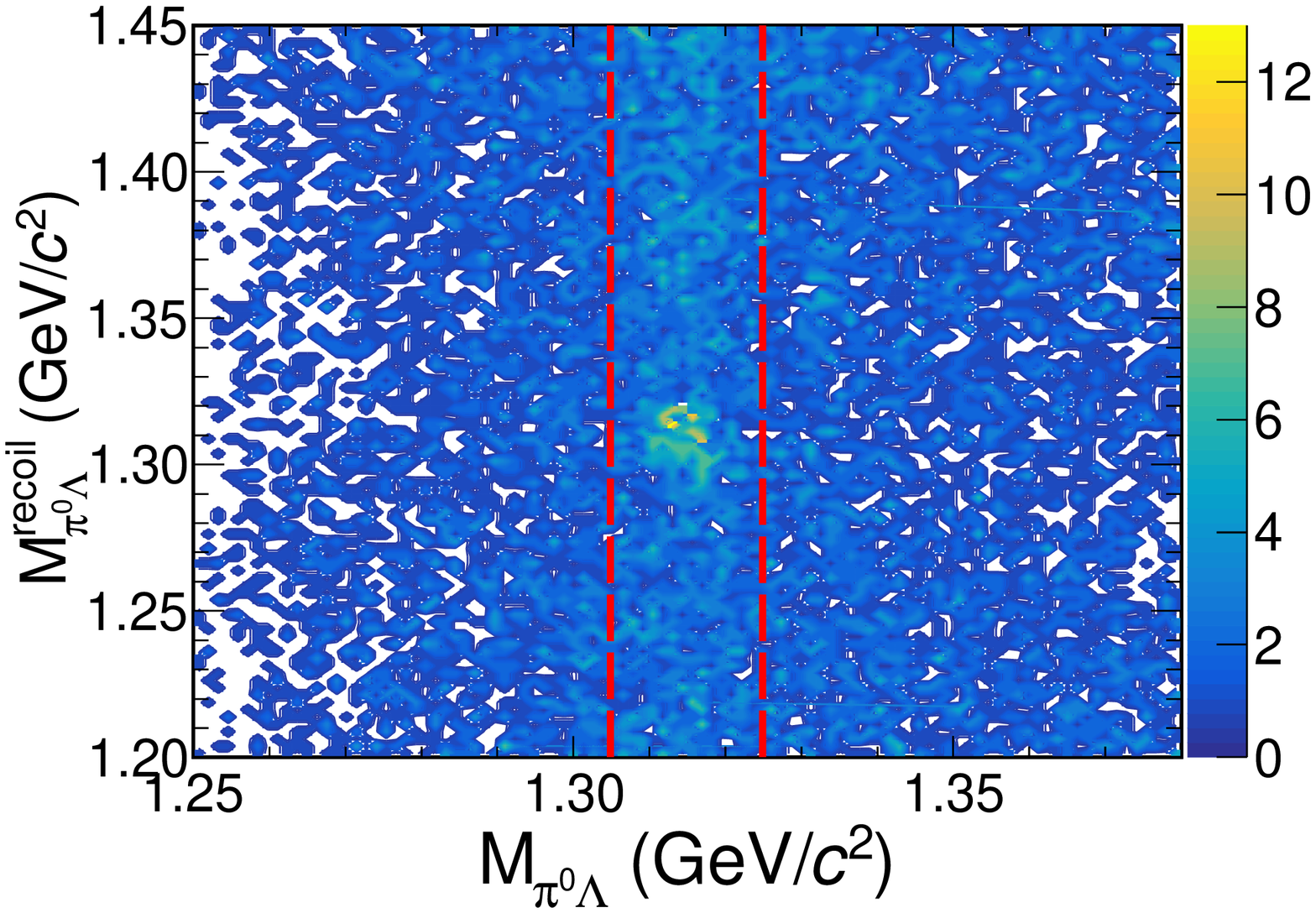}\\
\caption{Distribution of $M_{\pi^{0}\Lambda}$ versus $M^{\rm recoil}_{\pi^{0}\Lambda}$ for all energy points from data. The dashed lines denote the $\Xi^{0}$ signal region.}
\label{scatterplot}
\ecl
\end{figurehere}
\section{Extraction of signal yields}
The signal yield for the process $\EE\ar\XXN$  at each energy point is determined
by performing an extended unbinned maximum-likelihood
fit to the
$M^{\rm recoil}_{\pi^{0}\Lambda}$ spectrum in the range from 1.20
GeV/$c^{2}$ to 1.45 GeV/$c^{2}$.
In the fit, the signal shape for the process
$\EE\ar\XXN$  at each energy point is represented by the
simulated MC shape.
 After applying the same event selection
on the inclusive MC samples at each CM energy, it is found that some
background is present at each energy point, mainly coming from $\EE\ar\pi\Sigma\bar\Sigma$ events. 
These events distribute smoothly in the region of interest and can be described by a second-order polynomial function, while for two energy points near the $\XXN$ mass threshold at 
2.644 and 2.646~GeV, the background is modeled by an Argus function~\cite{argus}.
The signal significances for the energy points 2.644, 2.646, 2.700 and 2.950 GeV are found
 to be below three standard deviations. 
For these energy points, the upper limits for $\EE\ar\XXN$ production are calculated at the 90\% confidence
level (C.L.) based on the profile likelihood method incorporating systematic uncertainties~\cite{Stenson}.
Figure~\ref{mass_pi0lambda} shows the fit to the distribution of $M^{\rm recoil}_{\pi^{0}\Lambda}$ for the $\EE\ar\XXN$ process at each energy point  after applying the 
event selection described above.
Note that single-baryon-tag method means that in a small fraction of events both the $\Xi^0$ and $\bar{\Xi}^0$ hyperons are reconstructed, which means that the recoil-mass fit introduces double counting into the analysis. 
A correction factor of 1.08 is applied to the statistical uncertainty to account for this effect~\cite{Ablikim:2020rwi, Ablikim:2019wjs} based on MC simulation,
while for two energy points near the $\XXN$ mass threshold at 2.644 and 2.646~GeV, the correction factor is estimated to be 1.03. 
The number of observed events is summarized in Table~\ref{results:BCS:EFF}.  

\section{Determination of Born cross section}
The Born cross section for $\EE\ar\XXN$ is calculated by
\begin{equation}
\sigma^{B}(s) =\frac{N_{\rm obs}}{2{\cal{L}}(1 + \delta)\frac{1}{|1 - \Pi|^2}\epsilon{\cal B}},
\end{equation}
where $N_{\rm obs}$ is the number of the observed signal events,
${\cal{L}}$ is the integrated luminosity related to the CM energy, the factor 2
 is required for the inclusion  of the $c.c.$ mode,
$(1 + \delta)$ is the ISR correction factor~\cite{Jadach:2000ir}, $\frac{1}{|1 - \Pi|^2}$ is the vacuum
polarization (VP) correction factor~\cite{Actis:2010gg}, $\epsilon$ is the detection
efficiency and ${\cal B}$ denotes the product of the known branching fractions for the decays $\Xi^{0}\ar\pi^{0}\Lambda$, $\Lambda\ar p\pi^{-}$ and $\pi^{0}\ar\gamma\gamma$, respectively~\cite{PDG2020}.  The ISR correction factor is obtained using the
QED calculation as described in Ref.~\cite{Kuraev:1985hb} and taking Eq.~(\ref{BCS})
used to fit the cross section measured in this analysis
parameterized as input.  The measured cross
sections are summarized in Table~\ref{results:BCS:EFF}.  
\begin{figurehere}
\bcl
\subfigure{\includegraphics[width=0.5\textwidth]{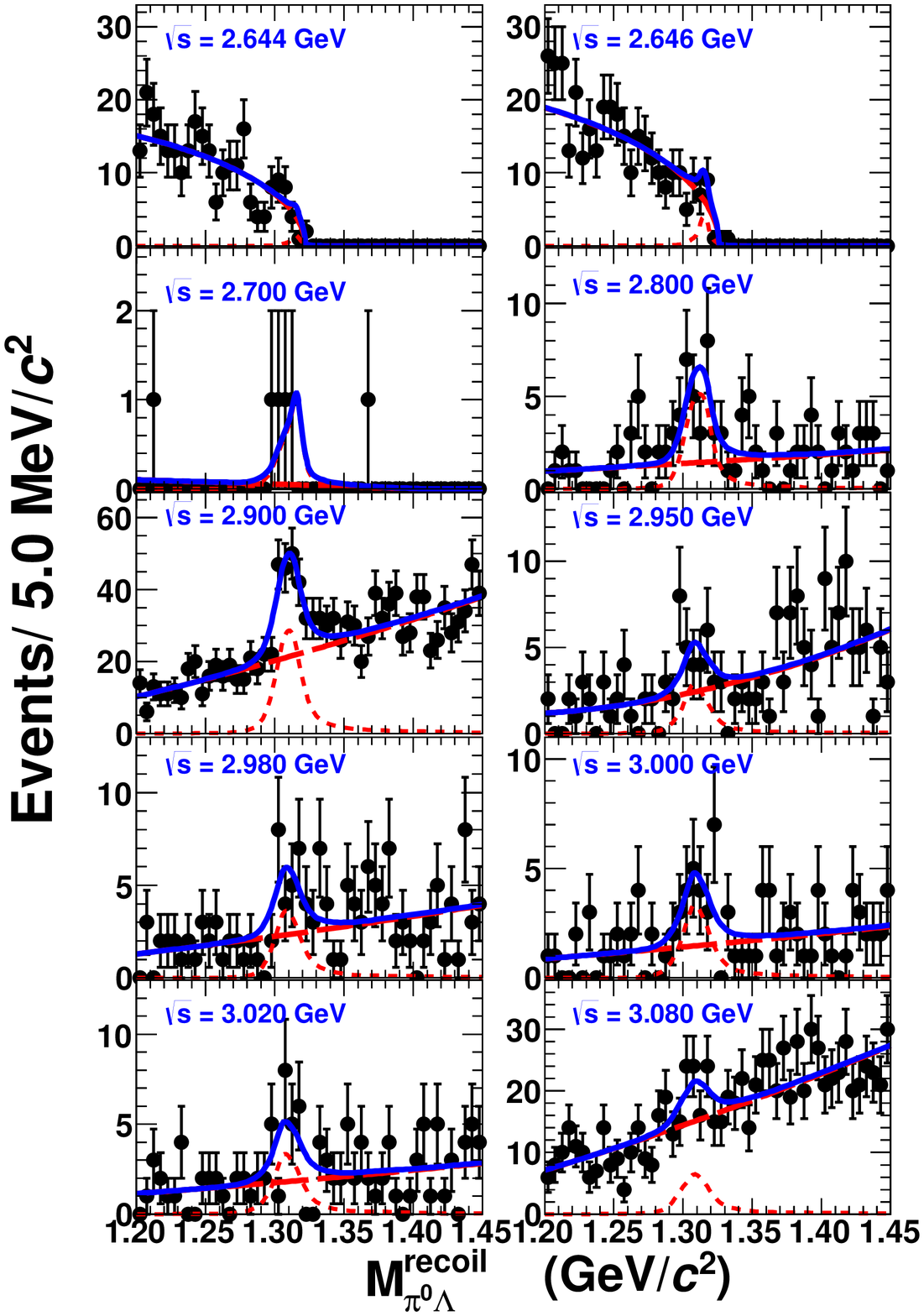}}
\caption{Fit to the recoil mass spectra of $\pi^{0}\Lambda$ for each energy point. Dots with error bars show data. The short-dashed line represents the signal shape and the long-dashed line signifies background.}
\label{mass_pi0lambda}
\ecl
\end{figurehere}
\btbl[!htb]
\caption{\small
The measured Born cross sections $\sigma^{B}$ and effective form factors $|G_{\rm eff}(s)|$ for $\EE\ar\XXN$  at ten CM energy points. Also listed at each point is the integrated luminosity ${\cal L}$,  the VP correction factor $\frac{1}{|1 - \prod|^{2}}$, the ISR correction factor $1 + \delta$ ,  the detection efficiency $\epsilon$, the number of the observed signal events $N_{obs}(N^{UP})$, and the signal significance $S(\sigma)$, taking account of both statistical and systematic uncertainties. 
The values between the parentheses for $\sigma^{B}$ and $|G_{\rm eff}(s)|$ are the corresponding upper limits at 90\% C.L.
}
\bcl
\doublerulesep 2pt
\scalebox{0.73}[0.73]
{
\begin{tabular}{c r@{.}l cc r@{ $\pm$ }l r@{ $\pm$ }l r@{ $\pm$ }c@{ $\pm$ }l r@{ $\pm$ }c@{ $\pm$ } l c} \hline\hline
$\sqrt{s}$ GeV  &\multicolumn{2}{c}{${\cal L}$ (pb$^{-1}$)} &$\frac{1}{|1 - \prod|^{2}}$  &$1 + \delta$ &\multicolumn{2}{c}{$\epsilon$ $(\%)$ }  &\multicolumn{2}{c}{$N_{obs}(N^{UP})$}    &\multicolumn{3}{c}{$\sigma^{B}$ (pb)} &\multicolumn{3}{c}{$|G_{\rm eff}(s)|$ $\times 10^{-3}$}     &$S(\sigma)$ \\ \hline
2.644   &33&72     &1.04     &0.75     &1.60 &0.03     &3.0 &2.8 ($< 7.9$)      &5.7  &5.4  &1.1 ($< 15.0$)     &54.0  &26.0 &5.0 ($< 88.0$)   &2.0 \\
2.646   &34&00     &1.04     &0.76     &1.71 &0.03     &9.3 &5.1 ($< 18.0$)     &16.1 &9.4  &2.8 ($< 31.2$)     &88.0  &26.0 &7.0 ($< 120.0$)  &2.0\\
2.700   &1&03      &1.04     &0.85     &5.85 &0.05     &3.8 &2.4 ($< 7.8$)      &56.8 &35.9 &3.1 ($< 116.5$)    &120.0 &38.0 &3.0 ($< 170.0$)  &2.1\\
2.800   &4&78      &1.04     &0.90     &10.38 &0.08    &27.8 &7.9               &47.6 &13.5 &2.6                &93.0  &13.0 &3.0              &5.0\\
2.900   &105&23    &1.03     &0.92     &12.08 &0.08    &160.0 &23.4             &10.5 &1.5  &0.6                &41.0  &3.0  &3.0              &8.2\\
2.950   &15&94     &1.03     &0.93     &12.09 &0.08    &16.7 &9.2 ($< 28.8$)    &7.2  &4.0  &0.4 ($< 12.4$)     &34.0  &9.3  &1.0 ($< 44.0$)   &2.7 \\
2.981   &16&07     &1.02     &0.94     &12.80 &0.08    &20.5 &8.2               &8.2  &3.3  &0.4                &36.0  &7.2  &1.0              &3.2\\
3.000   &15&88     &1.02     &0.94     &12.67 &0.08    &19.9 &7.3               &8.2  &3.0  &0.4                &36.0  &6.5  &1.0              &3.6\\ 
3.020   &17&29     &1.01     &0.95     &12.80 &0.08    &20.3 &7.6               &7.6  &2.8  &0.4                &34.0  &6.4  &1.0              &3.6\\ 
3.080   &263&70    &0.91     &0.97     &12.60 &0.08    &54.0 &22.2              &1.5  &0.6  &0.1                &15.0  &3.0  &0.4              &3.0
\\ \hline\hline
\end{tabular}}
\label{results:BCS:EFF}
\ecl
\etbl
\section{Determination of effective form factor}
Assuming  that spin-1/2 baryon pair production is dominated by one-photon exchange, the Born cross-section for the process $\EE\ar\XXN$ can be parameterized in terms of the two EMFFs $G_{E}(s)$ and $G_{M}(s)$~\cite{Cabibbo:1961sz} as follows:
\begin{equation}\label{FF02}
\sigma^{B}(s) = \frac{4\pi\alpha^{2}\beta}{3s} [|G_{M}(s)|^{2} + \frac{1}{2\tau}|G_{E}(s)|^{2}],
\end{equation}
where $\alpha$ is the fine structure constant, the variable $\beta =\sqrt{1-1/\tau}$ is 
 the $\BB$ velocity, $\tau = s/4m^{2}_{\Xi^0}$, $m_{\Xi^0}$ the $\Xi^0$ mass, and $s$ is the square of the measured CM energy. 
The effective form factor  is defined as a combination of the EMFFs
\begin{equation}\label{FF04}
|G_{\rm eff}(s)|= \sqrt{
\frac{2\tau|G_{M}(s)|^{2} + |G_{E}(s)|^{2}}{2\tau +1}},
\end{equation}
and, through substitution of 
Eq.~(\ref{FF02}) into Eq.~(\ref{FF04}), is proportional to the square root of the Born cross-section:
\begin{equation}
|G_{\rm eff}(s)|  = \sqrt{\frac{3s\sigma^{B}}{4\pi\alpha^2\beta\left(1+\frac{2m^{2}_{\Xi^{0}}}{s}\right)}}.
\end{equation}
The measured values of the effective form factors are summarized in Table~\ref{results:BCS:EFF}.  

\section{Fit to Born cross section}
A least-$\chi^2$ method~\cite{ZHUYS} is used to fit the Born cross-section for the process $\EE\ar\XXN$  with the assumption of two alternative functions. The first of these is 
a perturbative-QCD (pQCD) driven energy power function~\cite{Pacetti}, 
\begin{equation}
\sigma^{B} =\frac{c_0\cdot\beta}{(\sqrt{s}-c_1)^{10}},
\label{BCS}
\end{equation}
where $c_0$ is the normalization, $c_1$ is the mean effect of a set of possible intermediate states.
This model has been applied successfully in studies of $\EE\ar\llb$~\cite{LLbar}, $\Sigma^{\pm}\bar\Sigma^{\mp}$~\cite{Ablikim:2020kqp}, and $\XXB$~\cite{Ablikim:2020rwi} production.
The fit returns  $c_0 =$ (20.3 $\pm$ 9.6) pb $\cdot$ GeV$^{-10}$ and $c_1 =$ (1.52 $\pm$ 0.13) GeV, where the uncertainty includes both statistical and systematic contributions. Figure~\ref{Fig:XiXi::CS::Line-shape::Fit_BW} shows the fit result with quality $\chi^2/ndof =19.9/8.0$, where the number of freedom degree ($ndof$) is calculated by subtracting the number of free parameters in the fit from the total number of energy points.
The second function is a coherent sum of a pQCD-driven energy power function plus a Breit-Wigner (BW) function to test the resonance reported in the charged mode~\cite{Ablikim:2020rwi}
\begin{equation}\label{func11}
\sigma^{B}(\sqrt{s}) =\left|\sqrt{\frac{c_0\cdot\beta}{(\sqrt{s}-c_1)^{10}}} + e^{i\phi}BW(\sqrt{s})\sqrt{\frac{P(\sqrt{s})}{P(M)}}\right|^{2}.
\end{equation}
Here
the mass and width are fixed to the values of the charged mode~\cite{Ablikim:2020rwi}, 
$\phi$ is a relative phase between the BW function
\begin{equation}\label{func12}
BW(\sqrt{s}) =\frac{\sqrt{12\pi\Gamma_{ee}{\cal{B}}\Gamma}}{s-M^{2}+iM\Gamma},
\end{equation} 
and the power function and $P(\sqrt{s})$ is the two-body phase space factor.
In Eq.~(\ref{func12}), $\Gamma_{ee}$ is the electronic partial width, 
$M$ and $\Gamma$ are the mass and width of BW, respectively, and  ${\cal{B}}$ is the branching fraction.
The
significance for the resonance described by the BW is estimated to be 2.0$\sigma$ including the systematic uncertainty, which receives contributions from the Born cross section for this measurement and the uncertainty in the knowledge  of the  mass and width of the resonance from  previous studies~\cite{Ablikim:2020rwi}.  The fit has a quality of $\chi^2/ndof =10.7/4.0$ and is shown in Figure~\ref{Fig:XiXi::CS::Line-shape::Fit_BW}.  
The comparison of fit quality for both models is fair.
Note that neither fit model describes the point near 2.8 GeV well, which may be due to statistical fluctuation or the contribution of one or more additional unknown resonances.
A Bayesian approach~\cite{Bayesian} gives the upper limit on the product of the electronic partial width and the branching fractions for this possible resonance decaying to the $\XXN$  to be $\Gamma_{ee}{\cal{B}}$ $<$ 0.3 eV at the 90\% C.L. , after taking into account the systematic uncertainties described in the following. 
\begin{figurehere}
\bcl
\includegraphics[width=0.4\textwidth]{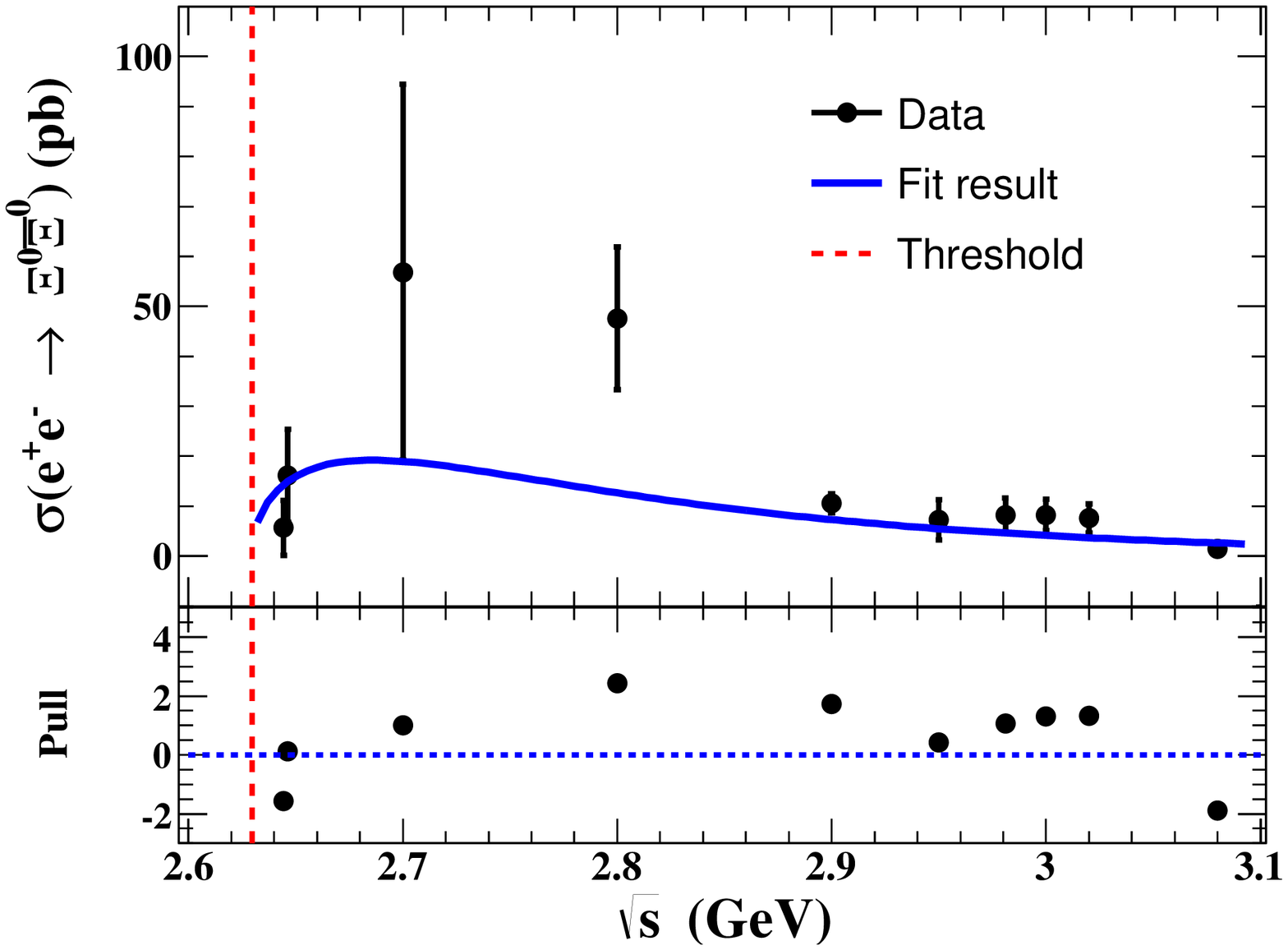}
\includegraphics[width=0.4\textwidth]{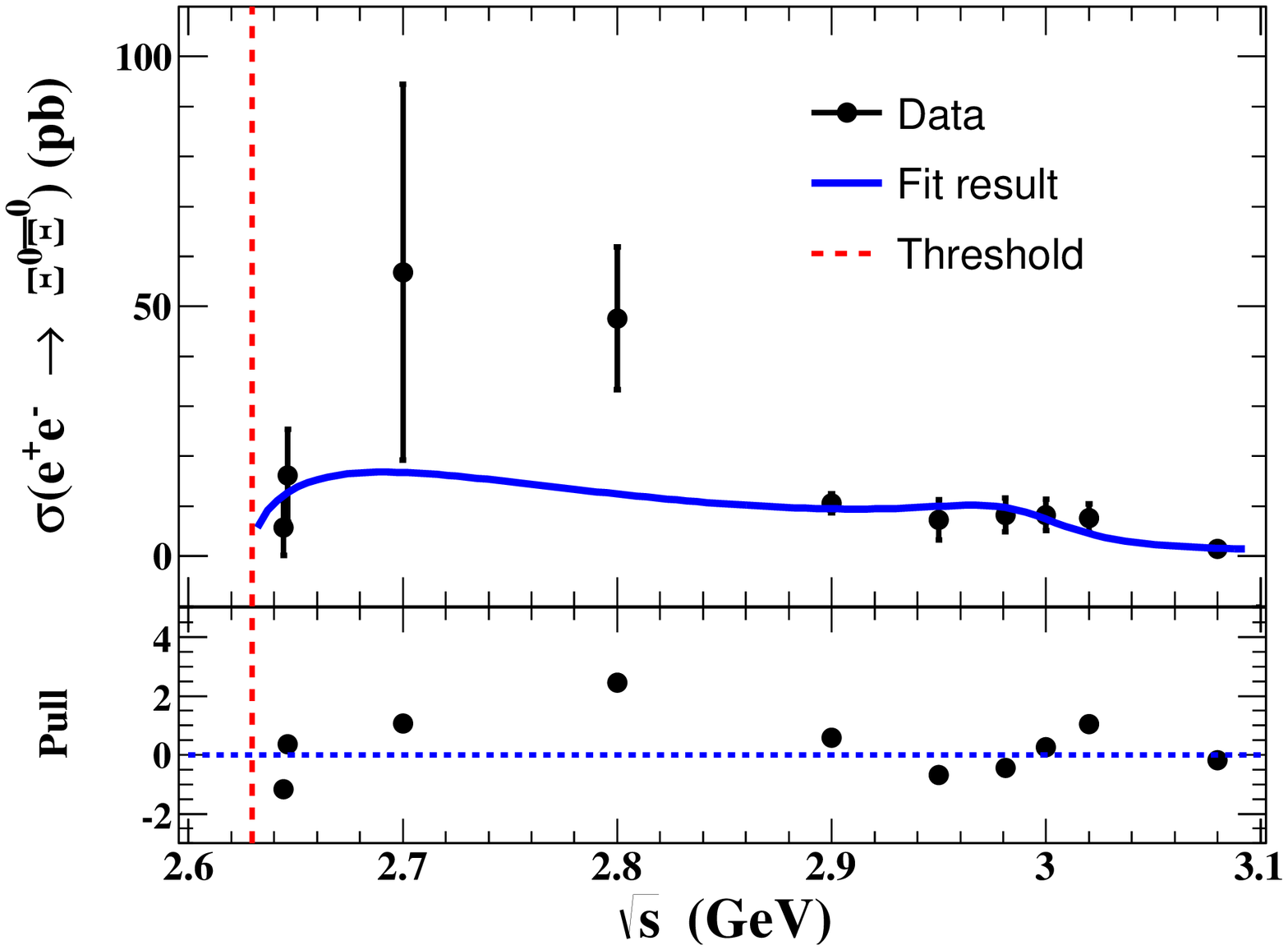}
\caption{Fit to the Born cross section of $\EE\ar\XXN$ at CM energies between 2.644 and 3.080 GeV for the first assumed function (top) and second assumed function (bottom) discussed in the main text. Dots with error bars show the Born cross section  including both statistical and systematic uncertainties. The blue solid line is the fit result. The vertical dashed line represents the production threshold for $\XXN$ pairs. The bottom panel of each plot gives the pull distribution of fit.
}
\label{Fig:XiXi::CS::Line-shape::Fit_BW}
\ecl
\end{figurehere}

\section{Systematic uncertainty}
\label{sec:syst_err}
Several sources of systematic uncertainties are considered on the Born cross section measurement. They include the $\Xi$ reconstruction efficiency,
 the fit range, the background shape, the mass resolution and the fit models for the cross section.
Knowledge of  the decay branching fractions of intermediate states and the luminosity measurement give additional contributions.
All of the systematic uncertainties are discussed in detail below.
\begin{enumerate}
   \item The understanding of the $\Xi^{0}$ reconstruction efficiency, which in turn depends on the photon-selection efficiency, the $\pi^{0}$ reconstruction efficiency, the
  $\Lambda$ reconstruction efficiency,  the requirements on the mass of the $\Lambda$ and $\Xi^{0}$, and the  decay length of the $\Lambda$ candidates,
   is studied with a control sample of $\jpsi\ar\XXN$ events  
via single- and double-tag methods. The selection criteria for the charged tracks, and the reconstruction of $\Lambda$ and $\Xi^{0}$ candidates are exactly the same as those described in Sec.~\ref{sec:evt_sel}. The $\Xi^{0}$ reconstruction efficiency is defined as the ratio of the number of double-tag $\XXN$ events to that from the single tags.
For studying the reconstruction efficiency for two energy points near threshold, care is taken to ensure that the momenta of the two pions in the final state of the control sample is within the same range as those of the signal decays.
A detailed description of the method can be found in Ref.~\cite{Ablikim:2006aw}. The difference in the $\Xi^{0}$ reconstruction efficiency between the data and MC samples is assigned as the uncertainty.

\item The uncertainty due to the fitting range is estimated by shifting the fitting range of $M^{\rm recoil}_{\pi^{0}\Lambda}$ by 20 MeV/$c^{2}$. The difference in yields is  taken as the systematic uncertainty of the fit range.

\item The uncertainties due to the background shape arise mainly from the level of the wrong-combination background and the form of the polynomial function used. For the wrong-combination background, 
there are transition $\pi^{0}$’s with similar momenta in both the baryon and anti-baryon decay chains within the signal events. Incorrect use of these in the $\Xi^{0}$ reconstruction leads to the wrong-combination background. The impact of this component on the background shape is studied by studying how the shape changes as the association criteria for the neutral pions are varied in the MC.
The uncertainty associated with the polynomial function  is estimated by changing from a second order to a third order function.
The observed difference in the yields is taken as the systematic uncertainty due to the background shape.

\item The uncertainty associated with knowledge of  the mass resolution is estimated by changing the nominal PDF to the MC-simulated shape
 convolved with Gaussian function whose mean and width are free parameters. The difference in yields is taken as the systematic uncertainty due to the mass resolution.   

\item 
In this analysis, the detection efficiency for each energy point is determined from MC simulated with the phase space model, which may not describe the angular distribution well.
Thus, a joint angular distribution formalism~\cite{Adlarson:2019jtw} combined with the measured parameters in $J/\psi\ar\XXN$ decay is employed to validate the detection efficiency, 
where the angular distribution parameter is set to unity which is its maximally allowed value.
The difference in the efficiency for both models is taken as the systematic uncertainty associated with the angular distribution.

 \item The uncertainty of the fit model used to evaluate the line shape of Born cross section incorporating the effect of ISR factor 
is estimated
 by floating and fixing the index of the pQCD energy power function. 
    The resulting change in selection efficiency is taken as the systematic uncertainty due to the fitting model of line shape. 
    \item The uncertainties associated with the branching fractions of the intermediate states $\Xi^{0}$ and $\Lambda$ are taken from the PDG~\cite{PDG2020}.
    \item The luminosity at all energy points is measured using the Bhabha events, with an uncertainty of
    about 1.0\%~\cite{zhangbinxin}.
\end{enumerate}

The various systematic uncertainties on the cross-section measurements are summarized in Table~\ref{systematic}, where the values in parentheses represent the corresponding values for the two energy points near threshold. Assuming all sources to be independent, the total systematic uncertainty is obtained by summing over the individual contributions in quadrature.
\begin{tablehere}
\caption{\small Systematic uncertainty on the Born cross section measurement, with the values in parentheses indicating  the value near threshold (in \%).}
\bcl
\begin{tabular}{ccc}  \hline \hline
Source                                                       &Value\\ \hline
$\Xi^{0}$ reconstruction                            &2.7 (14.8)\\
Fitting range                                              &2.5 \\
Background shape                                   &3.3\\
Mass resolution                                       &1.0\\
Angular distribution                                  &1.0\\
Line shape                                               &1.1\\  
Intermediate states                                  &0.8 \\ 
Luminosity                                               &1.0\\
Total                                                        &5.4 (15.5)\\ \hline \hline
\end{tabular}
\label{systematic}
\ecl
\end{tablehere}
\section{Discussion and conclusion}
\label{sec:conclusion}
Born cross sections and effective form factors for the exclusive process $\EE\ar\XXN$ are measured by means of a single baryon tag method for the first time. The $e^+e^-$ collision data are collected at ten center-of-mass energies
between 2.644 and 3.080 GeV  by the BESIII detector
 at BEPCII, and correspond to a total integrated luminosity of about 500 pb$^{-1}$ . 
The measured Born cross sections are described adequately with a pQCD-driven energy power function, which tend to zero around 2.64 GeV and so do not exhibit any obvious threshold enhancement. 
Allowing for a resonance at around 3.0~GeV, as reported in  Ref.~\cite{Ablikim:2020rwi}, gives a signal with a significance of 2.0$\sigma$, including both statistical and systematic uncertainties.
The upper limit on the product of the electronic partial width and the branching fractions for this resonance decaying to the $\XXN$ final state is  estimated to be $\Gamma_{ee}{\cal{B}} < $ 0.3 eV at 90\% C.L..  The fit models considered do not describe the data well at 2.8 GeV, which is behavior that warrants closer study  when larger data sets become available.
The effective form factor is determined at each energy point, but the samples are not large enough to allow separate measurements of $G_E(s)$ and $G_M(s)$.
A comparison of the measured Born cross sections
between the charged mode $\EE\ar\XXB$~\cite{Ablikim:2020rwi} and the neutral mode $\EE\ar\XXN$, as shown in 
Fig.~\ref{Fig:XiXi::CS::Line-shape:Comparison}, shows consistent behavior with the current level of experimental precision.
\begin{figurehere}
\bcl
\includegraphics[width=0.4\textwidth]{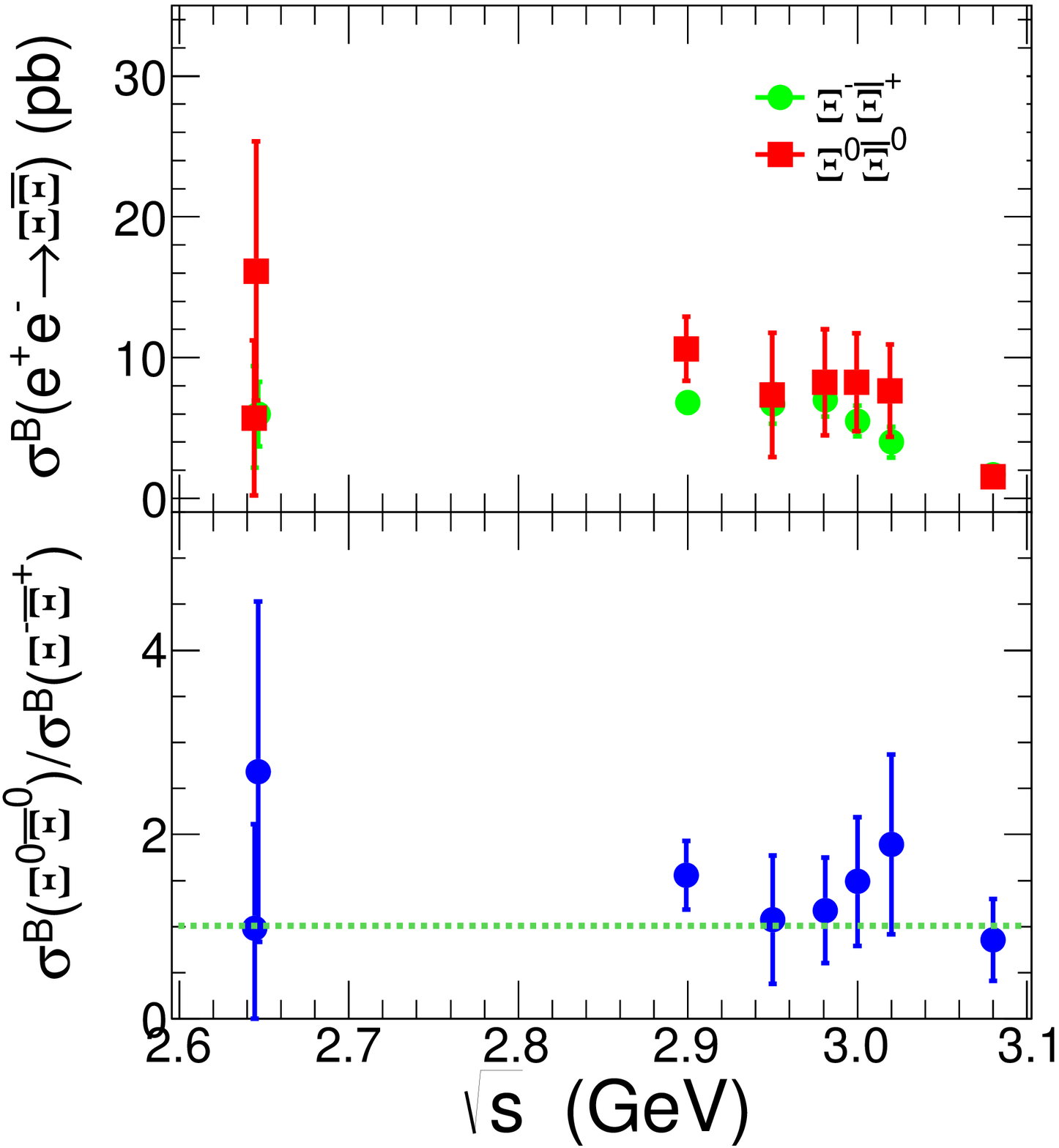}
\caption{
Comparison of Born cross sections between the charged mode~\cite{Ablikim:2020rwi} and neutral mode (top) and the ratio of Born cross sections between both modes bottom for eight energy points from 2.6444 to 3.0800 GeV, where the uncertainties include both statistical and systematic contributions.}
\label{Fig:XiXi::CS::Line-shape:Comparison}
\ecl
\end{figurehere}
The ratio of  Born cross sections for both modes is within 1$\sigma$ of the expectation of isospin symmetry.
These results  provide more information for understanding the insight into the nature of  hyperon pair production in $\EE$ annihilation near threshold. 

\section{Acknowledgements}

The BESIII collaboration thanks the staff of BEPCII and the IHEP computing center for their strong support. This work is supported in part by National Key Research and Development Program of China under Contracts Nos. 2020YFA0406400, 2020YFA0406403, 2020YFA0406300; National Natural Science Foundation of China (NSFC) under Contracts Nos. 11625523, 11635010, 11735014, 11822506, 11835012, 11905236, 11935015, 11935016, 11935018, 11961141012, 12022510, 12035009, 12035013, 12047501, 12061131003, 12075107; the Chinese Academy of Sciences (CAS) Large-Scale Scientific Facility Program; Joint Large-Scale Scientific Facility Funds of the NSFC and CAS under Contracts Nos. U1732263, U1832207; CAS Key Research Program of Frontier Sciences under Contract No. QYZDJ-SSW-SLH040; 100 Talents Program of CAS; INPAC and Shanghai Key Laboratory for Particle Physics and Cosmology; ERC under Contract No. 758462; European Union Horizon 2020 research and innovation programme under Contract No. Marie Sklodowska-Curie grant agreement No 894790; German Research Foundation DFG under Contracts Nos. 443159800, Collaborative Research Center CRC 1044, FOR 2359, FOR 2359, GRK 214; Istituto Nazionale di Fisica Nucleare, Italy; Ministry of Development of Turkey under Contract No. DPT2006K-120470; National Science and Technology fund; Olle Engkvist Foundation under Contract No. 200-0605; STFC (United Kingdom); The Knut and Alice Wallenberg Foundation (Sweden) under Contract No. 2016.0157; The Royal Society, UK under Contracts Nos. DH140054, DH160214; The Swedish Research Council; U. S. Department of Energy under Contracts Nos. DE-FG02-05ER41374, DE-SC-0012069.

\end{multicols}

\begin{thebibliography}{**}
\bbt{Delcourt:1979ed} B.~Delcourt {\it et al.}, Phys.\ Lett.\  {\bf B} 86 (1979) 395.
\bbt{Bisello:1983at}  DM2 Collaboration, D.~Bisello {\it et al.},  Nucl.\ Phys.\ B {\bf 224} (1983) 379; Z.\ Phys.\ C {\bf 48} (1990) 23.
\bbt{Antonelli:1993vz} A.~Antonelli {\it et al.},   
  Phys.\ Lett.\ B {\bf 313} (1993) 283; Phys.\ Lett.\ B {\bf 334} (1994) 431; Nucl.\ Phys.\ B {\bf 517} (1998) 3.
\bbt{ppbar} BABAR Collaboration, J.\ P.\ Lees {\it et al.}, Phys.\ Rev.\ D {\bf 87} (2013) 092005; Phys.\ Rev.\ D {\bf 88} (2013) 072009.
\bbt{ppbar02} PS170 Collaboration, G.~Bardin {\it et al.}, Nucl.\ Phys.\ B {\bf 411} (1994) 3.
\bbt{Akhmetshin:2015ifg} CMD-3 Collaboration
  R.~R.~Akhmetshin {\it et al.} ,  Phys.\ Lett.\ B {\bf 759} (2016) 634
\bbt{Pedlar:2005sj} CLEO Collaboration, T.~K.~Pedlar {\it et al.}, Phys.\ Rev.\ Lett.\  {\bf 95} (2005) 261803.
\bbt{ppbar03} BESIII Collaboration, M. Ablikim {\it et al.}, Phys.\ Rev.\ D {\bf 91} (2015) 112004; Phys.\ Rev.\ Lett.\  {\bf 124} (2020) 042001; Phys. Lett. B \textbf{817} (2021) 136328.
\bbt{nnbar} M.~N.~Achasov {\it et al.},  Phys.\ Rev.\ D {\bf 90} (2014) 112007.

\bibitem{Ablikim:2021eqa} BESIII Collaboration, M. Ablikim {\it et al.},
[arXiv:2103.12486 [hep-ex]].


\bibitem{Druzhinin:2019gpo}
  V.~P.~Druzhinin and S.~I.~Serednyakov, EPJ Web Conf.\  {\bf 212} (2019) 07007.
\bbt{LLbar}  BESIII Collaboration, M. Ablikim {\it et al.}, Phys.\ Rev.\ D {\bf 97} (2018) 032013. 
\bibitem{Pakhlova:2008vn} Belle Collaboration,
  G.~Pakhlova {\it et al.},  Phys.\ Rev.\ Lett.\  {\bf 101} (2008) 172001.
\bbt{LcLcbar}  BESIII Collaboration, M. Ablikim {\it et al.}, Phys.\ Rev.\ Lett.\  {\bf 120} (2018) 132001.
\bbt{theory1} O. D. Dalkarov, P. A. Khakhulin, and A. Y. Voronin, Nucl. Phys. A {\bf 833} (2010) 104.
\bbt{theory11} B. El-Bennich, M. Lacombe, B. Loiseau, and S. Wycech, Phys. Rev. C {\bf 79} (2009) 054001.  
\bbt{theory2} J.~Haidenbauer, H.~W.~Hammer, U.~G.~Meissner and A.~Sibirtsev, Phys.\ Lett.\ B {\bf 643} (2006) 29.
\bbt{theory22} R.~Baldini, S.~Pacetti, A.~Zallo, A.~Zichichi, Eur.~Phys.~J.~A {\bf 39} (2009) 315.
\bbt{Aubert} BABAR Collaboration, B.~Aubert {\it et al.}, Phys.\ Rev.\ D {\bf 76} (2007) 092006.
\bbt{DM2} DM2 Collaboration, D.~Bisello {\it et al.}, Z.\ Phys.\ C {\bf48} (1990) 23. 
\bbt{BESIII_hyperon} BESIII Collaboration, M.~Ablikim {\it et al.}, Phys.\ Rev.\ D {\bf 97} (2018) 032013;
Phys.\ Rev.\ Lett.\ {\bf 120} (2018) 132001; Chin.\ Phys.\ C {\bf 44} (2020) 040001.
\bbt{Ablikim:2020kqp} BESIII Collaboration, M.~Ablikim {\it et al.}, Phys. Lett. B \textbf{814} (2021) 136110.
\bbt{Ablikim:2020rwi}  BESIII Collaboration, M.~Ablikim {\it et al.}, Phys.\ Rev.\ D {\bf 103}, 012005 (2021).

\bbt{Denig:2012by} A.~Denig and G.~Salme, Prog. Part. Nucl. Phys. \textbf{68} (2013), 113-157.

\bbt{luminosity} With the same method (see Ref.~\cite{Ablikim:2017wlt} for more details), the preliminary luminosities for these energy points are determined
 to be a total of about 500 pb$^{-1}$ with an uncertainty of ~1\%.
\bbt{Ablikim:2017wlt} BESIII Collaboration,
M.~Ablikim \textit{et al.}, Chin. Phys. C \textbf{41} (2017) 063001; Chin. Phys. C \textbf{41} (2017) 013001; Chin. Phys. C \textbf{42} (2018) 023001.
\bbt{besiii} BESIII Collaboration, M. Ablikim {\it et al.}, Nucl. Instrum. Methods Phys. Res., Sect. A {\bf 614} (2010) 345.
\bbt{bepcii} Y. F. Wang, Int. J. Mod. Phys. A {\bf 21} (2006) 5371; C.~H.~Yu  {\it et al.}, Proceedings of IPAC2016, Busan, Korea, 2016, doi:10.18429/JACoW-IPAC2016-TUYA01.
\bibitem{Ablikim:2019hff} BESIII Collaboration, M. Ablikim {\it et al.},  Chin. Phys. C {\bf 44} (2020) 040001.
\bibitem{etof}
 X.~Li {\it et al.}, Radiat. Detect. Technol. Methods {\bf 1} (2017) 13;
 Y.~X.~Guo {\it et al.}, Radiat. Detect. Technol. Methods {\bf 1} (2017) 15;
 P.~Cao {\it et al.}, Nucl.\ Instrum.\ Meth.\ A {\bf 953} (2020) 163053.
\bbt{conexc} R.~G.~Ping {\it et al.} Chin. Phys. C {\bf 40} (2016) 113002.
\bbt{evtgen} D.~J.~Lange, Nucl.\ Instrum.\ Methods Phys.\ Res.\, Sect.\ A {\bf 462} (2001) 152; R.~G.~Ping {\it et al.} Chin.\ Phys.\ C {\bf 32} (2008) 599.
\bbt{PDG2020} Particle Data Group, P.~A.~Zyla {\it et al.},  Prog. Theor. Exp. Phys. {\bf 2020}, 083C01 (2020).
\bbt{geant4} GEANT4 Collaboration, S. Agostinelli {\it et al.}, Nucl. Instrum. Methods Phys. Res., Sect. A {\bf 506} (2003) 250; J.~Allison {\it et al.}, IEEE Trans.\ Nucl.\ Sci.\  {\bf 53} (2006) 270.
\bbt{XUM} M.~Xu  {\it et al.}, Chin.\ Phys.\ C {\bf33}, (2009) 428.
\bbt{argus} ARGUS Collaboration, H.~Albrecht {\it et al.}, Phys. Rev. Lett. {\bf 241} (1990)  278.
\bbt{Stenson} K.~Stenson, arXiv:physics/0605236;  J.~Lundberg, J.~Conrad, W.~Rolke and A.~Lopez, Comput.\ Phys.\ Commun.\ {\bf181} (2010) 683.
\bbt{Ablikim:2019wjs}  BESIII Collaboration, M.~Ablikim {\it et al.},  Phys.\ Rev.\ D {\bf 100} (2019) 051101.
\bibitem{Jadach:2000ir} 
  S.~Jadach, B.~F.~L.~Ward and Z.~Was,  Phys.\ Rev.\ D {\bf 63} (2001) 113009; R.~G.~Ping,  Chin.\ Phys.\ C {\bf 38} (2014) 083001.
\bibitem{Actis:2010gg} S.~Actis {\it et al.} Eur.\ Phys.\ J.\ C {\bf 66} (2010) 585.
\bbt{Kuraev:1985hb} E.~Kuraev and V.~S.~Fadin, Sov.\ J.\ Nucl.\ Phys.\ {\bf 41} (1985) 466.
\bbt{Cabibbo:1961sz} N.~Cabibbo and R.~Gatto, Phys.\ Rev.\  {\bf 124} (1961) 1577.

\bbt{ZHUYS}  Y.~S.~Zhu, Chin.\ Phys.\ C {\bf 32} (2008) 363;   G.~D' Agostini,
Bayesian Reasoning in Data Analysis: A Critical Introduction (World Scientific, New Jersey, USA (2003) 329 p.
\bbt{Pacetti}S.~Pacetti, R.~Baldini Ferroli and E.~Tomasi Gustafsson, Phys.\ Rep.\ {\bf 550} (2015) 1.
\bbt{Bayesian} Y.~S.~Zhu, Chin.\ Phys.\ C {\bf 32} (2008) 363; G.~D'Agostini, Bayesian reasoning in data analysis: A critical introduction,
  New Jersey, USA: World Scientific (2003) 329 p.
\bbt{Ablikim:2006aw} BESIII Collaboration, M.~Ablikim {\it et al.}, 
Phys.\ Rev.\ D {\bf 87} (2013) 032007; 
Phys.\ Rev.\ D {\bf 93} (2016) 072003 ;
Phys.\ Lett.\ B {\bf 770} (2017) 217;
Phys.\ Rev.\ D {\bf 100} (2019) 051101;
Phys.\ Rev.\ Lett.\ {\bf124} (2020) 032002.
\bibitem{Adlarson:2019jtw}
  P.~Adlarson and A.~Kupsc,  Phys.\ Rev.\ D {\bf 100} (2019) 114005.
\bbt{zhangbinxin} BESIII Collaboration, M.~Ablikim {\it et al.}, Chin.\ Phys.\ C {\bf 41} (2017) 063001; Chin.\ Phys.\ C {\bf 41} (2017) 113001.
\end{thebibliography}
\end{document}